\documentclass[conference]{IEEEtran}

\ifCLASSINFOpdf
 
\else
 
\fi

\usepackage{array}
\usepackage{amsthm,amsmath,amssymb}
\usepackage{mathrsfs}
\usepackage{multirow}
\usepackage{siunitx}

\usepackage{booktabs}
\usepackage{graphicx}
\usepackage{enumerate}
\usepackage{cite}

\begin{document}
\newcommand{\ee}{\end{equation}}
\newcommand{\br}{{\mbox{\boldmath{$r$}}}}
\newcommand{\bp}{{\mbox{\boldmath{$p$}}}}
\newcommand{\bpi}{\mbox{\boldmath{ $\pi $}}}
\newcommand{\bn}{{\mbox{\boldmath{$n$}}}}
\newcommand{\balfa}{{\mbox{\boldmath{$\alpha$}}}}
\newcommand{\ba}{\mbox{\boldmath{$a $}}}
\newcommand{\bta}{\mbox{\boldmath{$\beta $}}}
\newcommand{\bg}{\mbox{\boldmath{$g $}}}
\newcommand{\bPsi}{\mbox{\boldmath{$\Psi $}}}
\newcommand{\bsigma}{\mbox{\boldmath{ $\Sigma $}}}
\newcommand{\bGamma}{{\bf \Gamma }}
\newcommand{\bA}{{\bf A }}
\newcommand{\bP}{{\bf P }}
\newcommand{\bX}{{\bf X }}
\newcommand{\bI}{{\bf I }}
\newcommand{\bR}{{\bf R }}
\newcommand{\bZ}{{\bf Z }}
\newcommand{\bz}{{\bf z }}
\newcommand{\bx}{{\mathbf{x}}}
\newcommand{\bM}{{\bf M}}
\newcommand{\bU}{{\bf U}}
\newcommand{\bD}{{\bf D}}
\newcommand{\bJ}{{\bf J}}
\newcommand{\bH}{{\bf H}}
\newcommand{\bK}{{\bf K}}
\newcommand{\bm}{{\bf m}}
\newcommand{\bN}{{\bf N}}
\newcommand{\bC}{{\bf C}}
\newcommand{\bL}{{\bf L}}
\newcommand{\bF}{{\bf F}}
\newcommand{\bv}{{\bf v}}
\newcommand{\bSigma}{{\bf \Sigma}}
\newcommand{\bS}{{\bf S}}
\newcommand{\bs}{{\bf s}}
\newcommand{\bO}{{\bf O}}
\newcommand{\bQ}{{\bf Q}}
\newcommand{\btr}{{\mbox{\boldmath{$tr$}}}}
\newcommand{\bNSCM}{{\bf NSCM}}
\newcommand{\barg}{{\bf arg}}
\newcommand{\bmax}{{\bf max}}
\newcommand{\test}{\mbox{$
	\begin{array}{c}
		\stackrel{ \stackrel{\textstyle H_1}{\textstyle >} } { \stackrel{\textstyle <}{\textstyle H_0} }
	\end{array}
	$}}
\newcommand{\tabincell}[2]{\begin{tabular}{@{}#1@{}}#2\end{tabular}}
\newtheorem{Def}{Definition}
\newtheorem{Pro}{Proposition}
\newtheorem{Lem}{Lemma}
\newtheorem{Exa}{Example}
\newtheorem{Rem}{Remark}
\newtheorem{Cor}{Corollary}
\renewcommand{\labelitemi}{$\bullet$}

\title{Trajectory PHD Filter with Unknown Detection\\ Profile and Clutter Rate}

\author{\IEEEauthorblockN{1\textsuperscript{st} Shaoxiu Wei}
	\IEEEauthorblockA{\textit{University of Electronic Science and} \\
	\textit{ Technology of China}\\
		Chengdu, China \\
		2293960749@qq.com}
	\and
	\IEEEauthorblockN{2\textsuperscript{nd} Boxiang Zhang}
	\IEEEauthorblockA{\textit{University of Electronic Science and} \\
	\textit{ Technology of China}\\
		Chengdu, China \\
		201811011921@std.uestc.edu.cn}
	\and
	\IEEEauthorblockN{3\textsuperscript{rd} Wei Yi*}
	\IEEEauthorblockA{\textit{University of Electronic Science and} \\
		\textit{ Technology of China}\\
		Chengdu, China \\
		kussoyi@gmail.com}
}

\maketitle

\IEEEpeerreviewmaketitle

\begin{abstract}
	In this paper, we derive the robust TPHD (R-TPHD) filter, which can adaptively learn the unknown detection profile history and clutter rate. The R-TPHD filter is derived by obtaining the best Poisson posterior density approximation over trajectories on hybrid and augmented state space by minimizing the Kullback-Leibler divergence (KLD). Because of the huge computational burden and the short-term stability of the detection profile, we also propose the R-TPHD filter with unknown detection profile only at current time as an approximation. The Beta-Gaussian mixture model is proposed for the implementation, which is referred to as the BG-R-TPHD filter and we also propose a $L$-scan approximation for the BG-R-TPHD filter, which possesses lower computational burden.
\end{abstract}

\begin{IEEEkeywords}
	Trajectory RFS, unknown detection profile history, unknown clutter process, multi-target tracking.
\end{IEEEkeywords}
\IEEEpeerreviewmaketitle

\section{Introduction}
\IEEEPARstart{T}{he} random finite set (RFS) approach becomes a research hot spot and developing trend of multi-target tracking (MTT) in radar recently, which aims to model the appearance and disappearance of targets, false detections and misdetections of measurement within the Bayesian framework \cite{RFS1}. 
\par The PHD filter \cite{PHD} is the most widely used filter based on RFS approaches and known for its low computational burden. The PHD filter propagates the first-order multi-target moments \cite{RFS-firstorder} through prediction and update step and considers a Poisson multi-target density \cite{Der-PHD}. There are several common implementations for the PHD filter, such as sequential Monte Carlo (SMC) \cite{SMC-PHD1} or Gaussian mixture (GM) \cite{PHD}. However, the PHD filter does not provide track information.

\par The trajectory PHD (TPHD) filter \cite{TPHD} addresses the intrinsic inability of the PHD filter to build tracks by using a set of trajectories as the posterior density information, instead of target labelling \cite{GLMB1,LMB}. Based on KLD minimization, the TPHD filter estimates trajectories by propagating the PHD of the Poisson multi-trajectory density. The Gaussian mixture model is proposed to obtain the closed-form solution of the TPHD filter, which is given as the GM-TPHD filter. Meanwhile, its $L$-scan approximation is used to achieve the fast implementation, which only updates the trajectories of the last $L$ time and keeps the rest unchanged.

\par In the MTT, knowledge of two uncertainty sources: the detection profile and clutter are also of significant importance. Due to the the time-varying nature of the detection probability and clutter
processes, it is inaccurate to apply model parameters chosen from training data
to the multi-target tracking filter at subsequent frames, while obtained from a set of measurements is an efficient method. Algorithms based on clutter generators 
\cite{UC1,UC2,UPD1} are widely used to learn the unknown clutter process \cite{UPD1}. For the unknown detection probability case, the augmented state model \cite{Robust0,Robust2,UPD1,UPD2} is proposed to denote the unknown detection profile in the PHD filter and the implementation is generally obtained by some distributions to describe the detection probability.

\par In this paper, we develop the TPHD filter to adapt to the unknown clutter rate and detection profile, deriving the robust TPHD (R-TPHD) filter by KLD minimization. The basic idea in this development is to augment the unknown detection probability history into the trajectory and use clutter generators to estimate the clutter rate. In addition, we also give an analytic form of the R-TPHD filter with unknown detection profile of a single frame as an approximation. In the implementation, we propose a Beta Gaussian mixture model \cite{UPD1} for the R-TPHD filter, the Beta Gaussian R-TPHD (BG-R-TPHD) filter and its $L$-scan approximation with lower computational burden. Simulation results demonstrate the BG-R-TPHD filter can achieve excellent performance.

\section{Background}
This section provides a brief review of the TPHD filter. Further details can be found in \cite{TPHD}.
\subsection{Notations}
The variable $X=(t,x^{1:i})$ denotes a single trajectory, where $t$ represents the birth time and $x^{1:i}=(x^1,...,x^i)$ denotes target states at each time step of the trajectory of length $i$. For a single trajectory state $X_k\in \mathbb{T}_k=\uplus_{(t,i)\in I_k}\{t\}\times\mathbb{R}^{in_x}$, where $\times$ denotes a Cartesian product, $I_k=\{(t,i):0\leq t\leq k~\text{and}~1\leq i\leq k-t+1\}$, $T_k$ denotes the space of trajectories and $\uplus$ denotes the disjoint union. The trajectory RFS is given as $\mathbf{X}_k=\{X_{1},...,X_{N^k}\}\subset \mathcal{F}(\mathbb T_k)$, where $\mathcal{F}(\mathbb{T}_k)$ is the respective collections of all finite subsets of $\mathbb T_k$. We use $D_k(X)$ to represent the PHD of posterior multi-trajectory density at time $k$. We use $\langle {a,b}\rangle$ denotes the inner product between two real valued functions $a$ and $b$, which equals to $\int {a( x )b(x)} dx$. The indicate function \cite{TPHD} is given as $1_X(Y)$.
\subsection{Bayesian Filtering Recursion for Trajectories}
Given the posterior multi-trajectory density $p_{k-1}$ at time $k-1$, the posterior density at time $k$ can be given by using the Bayes recursion
\begin{align}
	\ p_{k|k-1}\left(\mathbf{X}_k \right) =& \int {f\left( {\mathbf{X}_k |\mathbf{X}_{k-1} } \right)} {p_{k - 1}}\left( \mathbf{X}_{k-1} \right)\delta \mathbf{X}_{k-1} , \\
	\ p_k\left( \mathbf{X}_k \right) =& \frac{{{\ell_k}\left( {{Z_k}|\mathbf{X}_k} \right){p_{k|k - 1}}\left( \mathbf{X}_k \right)}}{{{\ell_k}\left( {{Z_k}} \right)}},
\end{align}
where $f( { \mathbf{X}_k | \mathbf{X}_{k-1} })$ denotes the transition density, $ p_{k|k - 1}(\mathbf{X}_k)$ denotes the predicted density, ${\ell_k}\left({Z_k}|\mathbf{X}\right)$ denotes the density of measurements of trajectories. Because the measurements are only based on the current target states, ${\ell_k}\left( {{Z_k}}|\mathbf{X}_k \right)$ can be also written as
\begin{equation}
	{\ell_k}\left( {{Z_k}|\mathbf{X}_k} \right) = {\ell_k}\left( {{Z_k}|{\tau _k}\left( \mathbf{X}_k \right)} \right),
\end{equation}
where $\tau_k(\mathbf X)$ denotes the corresponding multi-target state at the time $k$. The denominator of the update step of Bayes recursion can be written as
\begin{equation}
	{\ell_k}\left( {{Z_k}} \right) = \int {{\ell_k}\left( {{Z_k}|{\tau _k}\left( \mathbf{X}_k \right)} \right)} {p_{k|k - 1}}\left( \mathbf{X}_k \right)\delta \mathbf{X}_k,
\end{equation}
\subsection{The TPHD Filter}
\par The TPHD filter considers a Poisson probability density function (PDF). Thus, the posterior multi-trajectory density $p(\cdot)$ of a Poisson RFS can be written as
\begin{equation}
	p_k(\{X_1,...,X_{N^k}\})=e^{-\lambda_k}\lambda_k^n\sum_{j=1}^{N^k}\bar{p}_k(X_j),
\end{equation}
where $\bar{p}_k(\cdot)$ is a single trajectory density and $\lambda_k\ge0$. A Poisson PDF is characterized by its PHD $D_k(X) = \lambda_k\bar{p}_k(X)$ \cite{Der-PHD}. The clutter RFS is also given as a Poisson process of $\lambda_c$ with intensity $\bar c(\cdot)$. 
\par There are some assumptions,
\begin{itemize}
	\item [$\bullet$]
	At time $k$, the trajectories consist of surviving trajectories at time $k-1$ with surviving probability ${p_{S,k}}\left(  \cdot  \right)$, and the trajectories born at time $k$ with the PHD ${{\gamma_k }}( \cdot)$ of a Poisson density. The birth and the surviving RFSs are independent of each other.
	\item [$\bullet$]
	The trajectory RFS at time $k-1$ is Poisson and the clutter RFS is also Poisson and independent of measurement RFS.
\end{itemize} 
\par Based on KLD minimization, the TPHD filter will propagate the PHD of the Poisson multi-trajectory density. Given the posterior PHD $D_{k-1}(\underline X)$ at time $k-1$ and the transition density $f( {X|\underline X})$, where $\underline X$ denotes the trajectory state at time $k-1$, the prediction step of the TPHD filter is given as
\begin{equation}
	{D_{k|k - 1}}\left( X \right) = {\gamma _k}\left( X \right) + D_k^\zeta \left( X \right),
\end{equation}
where
\begin{align*}
	{\gamma _k}\left( X \right) &= \gamma \left( {k,{x^1}} \right),\\
	D_k^\zeta \left( X \right) &= {p_{S,k}}\left( {\underline X} \right)f\left( {X|\underline X} \right){D_{k - 1}}\left( {\underline X} \right).
\end{align*}
\par The update step of the TPHD filter is given by
\begin{equation}
	\begin{split}
		{D_k}\left( X \right) =& {D_{k|k - 1}}\left( X \right){q_{D,k}\left( x \right)}\\
		&+{D_{k|k - 1}}\left( X \right)p_{D,k}(x)\\
		&\times\sum\limits_{z \in {Z_k}} {\frac{{{l_k}(z|x)}}{{{\lambda _c}\bar c (z) + \left\langle {p_{D,k}\cdot{l_k}\left( {z| \cdot } \right),D_{k|k-1}^\tau } \right\rangle }}},
	\end{split}
\end{equation}
where
\begin{equation*}
	D_{k|k-1}^\tau \left( {x}\right)=\sum\limits_{t = 1}^k {\int {{D_{k|k - 1}}(t,{x^{1:k - t+1}})d{x^{1:k - t}}} },
\end{equation*}
which denotes the PHD of the prior target density at time $k$. In formula (7), $p_{D,k}(\cdot)$ denotes the detection probability, $q_{D,k}=1-p_{D,k}$, and $l_{k}(\cdot)$ denotes the measurement likelihood.
\section{The Proposed R-TPHD Filter}
In this section, the recursion of the R-TPHD Filter will be detailed, which can adaptively learn unknown detection profile and clutter rate, by using the hybrid and augmented state space model \cite{UPD1}. 
\subsection{The Hybrid and Augmented State Model}
Let the $\mathbb{T}_k=\uplus_{(t,i)\in I_k}\{t\}\times\mathbb{R}^{in_x}$ denote space of trajectories and $Y_k=\mathbb{O}^{i}$ to denote state space for its unknown detection probability history at time $k$, where $I_k=\{(t,i):0\leq t\leq k~\text{and}~1\leq i\leq k-t+1\}$ and $\mathbb{O}$ denotes the space in the interval of $[0,1]$. In addition, We use $\mathbb C_k$ to denote space of clutter and $\mathbb O_k$ to denote its corresponding detection probability. Thus, we define the hybrid and augmented trajectory space model as
$$ \mathbb U  = (\mathbb T  \times \mathbb Y) \uplus  (\mathbb C  \times \mathbb O) $$
where $\mathbb U$ represents the space of hybrid augmented trajectories ${\tilde X_c}$, which consists of augmented trajectories ${\tilde X}$ and augmented clutter state ${\tilde c}$. We define the density of hybrid augmented trajectory sets as hybrid augmented multi-trajectory density. For a single augmented trajectory ${\tilde X}= [ {X,A}] \in (\mathbb T \times \mathbb Y)$, $X=(t,x^{1:i}) \in \mathbb T$ denotes the trajectory and $A=a^{1:i} \in \mathbb Y$ denotes the detection probability history of this trajectory. On the other hand, at each historical moment of the augmented trajectory, we can obtain the augmented target state $\tilde{x}=(x,a)$, where $a\in\mathbb{O}$ denotes the detection probability of target and $x$ denotes the target state. We use $D_k(\tilde X)$ to denote the PHD of the augmented multi-trajectory density at time $k$. For the augmented clutter state ${\tilde c}= [ {c,o}] \in (\mathbb C  \times \mathbb O)$, $c$ denotes the clutter and $o\in\mathbb{O}$ denotes the detection probability of clutter. We use $D_k(\tilde c)$ to denote the PHD of the augmented clutter at time $k$.
\par Let the variable $\tilde p_{S,k}(\tilde X_c)$ denote the surviving probability and $\tilde f(\cdot|\cdot)$ denote the transition density. Meanwhile, it has detection probability $\tilde p_{D,k}(\tilde X_c)$ and the measurement likelihood $\tilde l_k(z|\tilde X_c)$. 
\par Besides, the clutter and trajectories are statistically independent and their augmented part $A$/$o$ and the state $X$/$c$ are also statistically independent. In addition, we consider that all clutter generators are of the same model, hence the clutter state can be ignored in recursions \cite{UPD1}. 
\par There are some points need to be noted. The surviving probability is given as
\begin{equation}
	{\tilde p_{S,k}}\left( {\tilde X_c} \right) =	\left\{
	\begin{array}{lr}
		{p_{S,k}}\left( x^i \right), &\tilde X_c\in\mathbb T\times \mathbb Y \\
		{p^c_{S,k}}, &\tilde X_c\in\mathbb C\times \mathbb O, 
	\end{array}
	\right.
\end{equation}
where the surviving probability of augmented trajectory $p_{S,k}(\tilde X)$ is only concerned about the target state at current time and the survival probability of the augmented clutter $p_{S,k}(\tilde c)=p^c_{S,k}$ is independent of clutter state. The transition density of hybrid augmented trajectory is given as
\begin{equation}
	\tilde f\left( {\tilde X_c|\underline{\tilde X}_c} \right) =	\left\{
	\begin{array}{lr}
		f\left( {{x^i}|{x^{i - 1}}} \right)f'\left( {a^i|a^{i-1}} \right), &\tilde X_c\in\mathbb T\times \mathbb Y \\
		&\underline{\tilde X}_c\in\mathbb T\times \mathbb Y\\
		f^c\left( c|\epsilon \right)f^o\left( {o|\tau} \right), 
		&\tilde X_c\in\mathbb C\times \mathbb O\\
		&\underline{\tilde X}_c\in\mathbb C\times \mathbb O. 
	\end{array}
\right.
\end{equation}
where $\underline{\tilde X}_c$ denotes the hybrid augmented trajectory at time $k-1$, $\epsilon$ denotes the clutter generator at time $k-1$ and $\tau$ the detection probability of clutter at time $k-1$.
\begin{equation}
	\tilde p_{D,k}(\tilde X_c) =	\left\{
	\begin{array}{lr}
		a^i, &\tilde X_c\in\mathbb T\times \mathbb Y \\
		o, &\tilde X_c\in\mathbb C\times \mathbb O. 
	\end{array}
	\right.
\end{equation}
\par The measurement likelihood of augmented trajectory ${l_k}( {z|\tilde X})$ is only concerned about target’s kinematic states and the measurement likelihood of augmented clutter is independent of the clutter state.
\begin{equation}
	{\tilde l_k}\left( {z|\tilde X_c} \right) =	\left\{
	\begin{array}{lr}
		{l_k}\left( {z|x^i} \right), &\tilde X_c\in\mathbb T\times \mathbb Y \\
		\bar c(z), &\tilde X_c\in\mathbb C\times \mathbb O. 
	\end{array}
	\right.
\end{equation}
\subsection{The R-TPHD Filter}
\par The recursion for the R-TPHD filter is based on the following assumptions:
\begin{itemize}
	\item[A1]The trajectories at current time consist of alive trajectories at last time and new births at current time, which are independent of
	each other. The augmented trajectory born at time $k$ is given as $\gamma_k ( {\tilde X})$.
	\item[A2]Each trajectory generates measurements independently of each other. Clutter is obtained by clutter generators.
\end{itemize}
\begin{Pro}\label{UTPHD_Pr}
	Given the posterior PHD $D_{k-1}$ at time $k-1$, the predicted PHD $D_{k|k-1}$ is obtained by
\end{Pro}
\begin{align}
	{D_{k|k - 1}}\left( {\tilde X} \right)&= {\gamma _k}\left( {\tilde X} \right) + D_k^\zeta \left( {\tilde X} \right),\\
	{D_{k|k - 1}}\left( {\tilde c} \right)&= {\gamma _k}\left( {\tilde c} \right) + p^c_{S,k}D_{k-1}\left( {\tilde c} \right),
\end{align}
where
\begin{align}
	{\gamma _k}\left( {\tilde X} \right) =&  \gamma \left( {k,{x^1},a^1} \right),\\
	D_k^\zeta \left( {\tilde X} \right) =&  {p_{S,k}}\left( {{x^{i - 1}}} \right)f\left( {{x^i}|{x^{i - 1}}} \right)
	\notag\\
	&\times f'\left( {a^i|a^{i-1}} \right){D_{k - 1}}\left( {t,{x^{1:i - 1}},a^{1:i - 1}} \right).
\end{align}
\par  In Proposition \ref{UTPHD_Pr}, it is required that $t + i - 1 = k$, since we only consider alive trajectories and $t \le k - 1$ denotes trajectories born before time $k$. Besides, the R-TPHD filter will retain past states of the detection profile history and trajectories in the PHD in Proposition \ref{UTPHD_Pr}, while traditional PHD filter \cite{PHD} only consider current states information, which will integrate all the past states. 
\begin{Pro}\label{UTPHD_Up}
	Given the predicted PHD $D_{k|k-1}$ at time $k$, the posterior PHD $D_{k}$ is obtained by
\end{Pro}
\begin{align}
	{D_k}\left( {\tilde X} \right) =& {D_{k|k - 1}}\left( {\tilde X} \right)(1-a^i)\notag\\
	& + {D_{k|k - 1}}\left( {\tilde X} \right)a^i\sum\limits_{z \in {Z_k}} {\frac{l_k(z|x^i)}{\Theta_k\left[z,\tilde c,\tilde x\right]}},\\
	{D_k}\left( {\tilde c} \right) =& {D_{k|k - 1}}\left( {\tilde c} \right)(1-o)\notag\\
	& + {D_{k|k - 1}}\left( {\tilde c} \right)o\sum\limits_{z \in {Z_k}}{\frac{\bar c_k(z)}{\Theta_k\left[z,\tilde c,\tilde x\right]}},
\end{align}
where
\begin{align}
	\Theta_k\left[z,\tilde c,\tilde x\right]=&\iint o\cdot \bar c_k(z)D_{k|k-1}(c,o)dcdo\\
	&+\iint a^i\cdot l_k(z|x^i)D^\tau_{k|k-1}(x^i,a^i)da^idx^i,\notag\\
	D_{k|k-1}^\tau(x^i,a^i)=&\sum\limits_{t = 1}^k\iint {D_{k|k - 1}(t,x^{1:k-t+1},a^{1:k-t+1})}\notag\\
	&{\times dx^{1:k-t}da^{1:k-t}}.
\end{align}
\par In Proposition \ref{UTPHD_Up}, $i=k-t+1$, $D_{k|k-1}^\tau (x^i,a^i)$ denotes the PHD of the prior density of augmented targets at time $k$ and $a^i$ denotes the detection profile at time $k$. The updated PHD contains information about the trajectory, clutter and detection profile history.
\section{The Beta Gaussian Mixture Implementation}
In this section, we will present a suboptimal implementation for the R-TPHD filter by using Beta-Gaussian mixture model \cite{UPD1}, which only consider unknown detection profile and clutter rate at current time. 
\subsection{Only Current Detection Profile}
In the update step of the R-TPHD filter, we update the whole detection profile history, trajectory and clutter generator. In
principle, we can correct the earlier estimation of trajectories by updating the detection profile history. However, due to the short-term stability of detection profile and a huge computational burden with an increasing trajectory, we simplify the R-TPHD filter to only consider the detection profile of trajectory at current time and the Beta distribution provides an efficient method to describe it \cite{UPD1}.
\par For the R-TPHD filter with unknown detection profile only at current time, the augmented trajectory can be simplified as $\tilde X=(t,x^{1:i},a^i)$. In this case, the prediction of augmented trajectory PHD can be simplified to
\begin{align}
	{D_{k|k - 1}}\left( {\tilde X} \right)&= {\gamma _k}\left( {\tilde X} \right) + p_{S,k}\left( {{x^{i - 1}}} \right)f\left( {{x^i}|{x^{i - 1}}} \right)
	\notag\\
	&\times\int f'\left( {a^i|a^{i-1}} \right){D_{k - 1}}\left( {t,{x^{1:i - 1}},a^{i-1}} \right)da^{i-1}.
\end{align}
Similarly, the PHD of the augmented target in will be changed into
\begin{align}
	D_{k|k-1}^\tau (x^i,a^i) =\sum\limits_{t = 1}^k {\int {{D_{k|k - 1}}(t,{x^{1:k - t+1}},a^{k-t+1})d{x^{1:k - t}}} }.
\end{align}
The rest of the derivation in the R-TPHD filter does not change. 
\subsection{The BG-R-TPHD Filter}
In this section, the Beta-Gaussian mixture model \cite{UPD1} is presented to obtain a suboptimal closed-form implementation for the R-TPHD filter. 
\par There are some notations given as follows. At time $k$, the notation ${\cal N}( {X;t,\widehat{m}^k,\widehat{P}^k}) = {\cal N}( {t,{x^{1:i}};\widehat{m}^k,\widehat{P}^k})1_{I}(t,i)$ denotes the Gaussian density of a trajectory which is born at time $t$ of length $i$. Its mean is $\widehat{m}^k \in \mathbb{R}^{i{n_x}}$ and the covariance is $\widehat{P}^k \in \mathbb{R}^{i{n_x}\times i{n_x}}$, where $n_x$ represents the number of dimension of targets' kinematics matrix. For a matrix $A$, We use ${A_{\left[ {n:m,s:t} \right]}}$ to represent its rows from time step $n$ to $m$ and columns from time step $s$ to $t$.
\par For a variable $y$ of Beta distribution \cite{UPD1}, its PDF is given as
\begin{equation}
	f(y)=\frac{y^{u-1}(1-y)^{v-1}}{\int_0^1{y^{u-1}(1-y)^{v-1}dy}},
\end{equation}
where $u > 1, v > 1$ and $\int_0^1{y^{u-1}(1-y)^{v-1}dy}$ denotes the Beta function $B(u,v)$. The Beta distribution has mean ${\mu _\beta } = \frac{u}{{u + v}}$ and variance $\sigma _\beta ^2 = \frac{{uv}}{{{{\left( {u + v} \right)}^2}\left( {u + v + 1} \right)}}$. For the Beta distribution $\beta \left( { \cdot ;u,v} \right)$, these are some identities as follows,
\begin{align}
	(1-y)\beta \left( { y ;u,v} \right) =& \frac{v}{u+v}\beta \left( { y ;u,v+1} \right),\\
	y\beta \left( { y ;u,v} \right) =& \frac{u}{u+v}\beta \left( { y;u+1,v} \right).
\end{align}
By using Beta distribution, the transition density for the detection profile is obtained by
\begin{align}
	f'\left( {a^i|a^{i-1}} \right) =&	\beta \left( { a ;u^{k|k-1},v^{k|k-1}} \right),\\
	f^o\left( {o|\tau} \right) =&	\beta \left( { o ;u_c^{k|k-1},v_c^{k|k-1}} \right).
\end{align}  
with $u,v$ comes from $a^{i-1}$ and $u_c,v_c$ comes from $\tau$. We use $a$ denote the latest state $a^{i}$ for simplicity. The detailed identities of Beta distribution can be found in \cite{UPD1}. There are some assumptions as follows 
\begin{itemize}
	\item[A7] The target’s kinematics and observation models are the linear Gaussian model
	\begin{align}
		f\left( {{x^i}|{x^{i - 1}}} \right)=&{\cal N}\left( {{x^i};F{x^{i - 1}},Q} \right),\\
		l\left( {z|x^i} \right)=&{\cal N}\left( {z;Hx^i,R} \right).
	\end{align}
	\item[A8] The detection probability only depends on the augmented part $a$ and $o$, and the surviving probability of trajectory and clutter are given as constants. 
	\item[A9] The PHD of birth trajectory and clutter density are given as
	\begin{align}
		{\gamma _k}\left( {\tilde X} \right) &= \sum\limits_{j = 1}^{J_\gamma ^k} {\omega _{\gamma ,j}^k\beta \left( {a;u_{\gamma ,j}^k,v_{\gamma ,j}^k} \right){\cal N}\left( {X;k,\widehat m_{\gamma ,j}^k,\widehat P_{\gamma ,j}^k} \right)}, \\
		{\gamma _k}\left( {\tilde c} \right)& = \sum\limits_{j = 1}^{J_\gamma ^{c,k}} {\omega _{c,\gamma ,j}^{k}\beta \left( {o;u_{c,\gamma ,j}^{k},v_{c,\gamma ,j}^{k}} \right)}, 
	\end{align}
	where $J_\gamma ^k$ denotes the number of birth trajectories, and $\omega_{\gamma,j}^k$ represents the corresponding weight with mean $\widehat m_{\gamma,j}^k \in \mathbb{R}^n_x$ and covariance $\widehat P_{\gamma,j}^k\in \mathbb{R}^{n_x \times n_x}$ , and $u_{\gamma ,j}^k,v_{\gamma ,j}^k$ denotes the corresponding factors of Beta distribution of the $j_{th}$ birth component. The PHD of birth clutter density is given in the same way as birth trajectory density, but ignore the state of clutter.
\end{itemize}
\begin{Pro}\label{UBGTPHD_Pr}
	The posterior PHD ${D_{k - 1}}( {\tilde X})$ and ${D_{k - 1}}( {\tilde c})$ at time $k-1$ can be respectively given by the
	Beta–Gaussian and Beta mixtures as follows
	\begin{align}
		{D_{k - 1}}\left( {\tilde X} \right) &= \sum\limits_{j = 1}^{{J^{k - 1}}} {\omega _j^{k - 1}\beta \left( {a;u_j^{k - 1},v_j^{k - 1}} \right)}\notag{\cal N}\left(X;t_j,\widehat{m}_j^k,\widehat{P}_j^k \right),\\
		{D_{k - 1}}\left( {\tilde c} \right) &= \sum\limits_{j = 1}^{{J_c^{k - 1}}} {\omega _{c,j}^{k - 1}\beta \left( {o;u_{c,j}^{k - 1},v_{c,j}^{k - 1}} \right)},\notag
	\end{align}
	where, at time $k-1$, the $j_{th}$ trajectory has length $i_j^{k-1}=k-t_j$, mean $\widehat m_j^{k-1}\in\mathbb{R}^{i_j^{k-1}\times n_x}$ and covariance $\widehat P_j^{k-1}\in \mathbb{R}^{i_j^{k-1}n_x \times i_j^{k-1}n_x }$.
	Then, the prior PHD $D_{k|k-1}(\tilde X)$ and $D_{k|k-1}(\tilde c)$ are given as, respectively
\end{Pro}
\begin{align}
	{D_{k|k - 1}}\left( {\tilde X} \right) =& \sum\limits_{j = 1}^{J_\gamma ^k} {\omega _{\gamma ,j}^k\beta \left( {a;u_{\gamma ,j}^k,v_{\gamma ,j}^k} \right)}\notag\\
	&{\times{\cal N}\left( {X;k,\widehat m_{\gamma ,j}^k,\widehat P_{\gamma ,j}^k} \right)} \\
	&+ {p_s}\sum\limits_{j = 1}^{{J^{k - 1}}} {\omega _j^{k - 1}\beta \left( {a;u_{S,j}^{k|k - 1},v_{S,j}^{k|k - 1}} \right)}\notag\\
	&{\times{\cal N}\left( {X;t_{j},\widehat{m} _{S,j}^{k|k-1},\widehat{P} _{S,j}^{k|k - 1}} \right)}\notag,\\
	{D_{k|k - 1}}\left( {\tilde c} \right) =& \sum\limits_{j = 1}^{J_\gamma ^{c,k}} {\omega _{c,\gamma ,j}^{k}\beta \left( {o;u_{c,\gamma ,j}^{k},v_{c,\gamma ,j}^{k}} \right)}\\
	&+ {p^c_s}\sum\limits_{j = 1}^{{J^{k - 1}}} {\omega _j^{c,k - 1}\beta \left( {o;u_{c,S,j}^{k|k - 1},v_{c,S,j}^{k|k - 1}} \right)},\notag
\end{align}
where
\begin{align}	
	u_{c,S,j}^{k|k - 1}=&u_{c,j}^{k - 1},
\end{align}
\begin{align}
		v_{c,S,j}^{k|k - 1} =& v_{c,j}^{k - 1},\\
	u_{S,j}^{k|k - 1}=& \left( {\frac{{\mu _j^{k|k - 1}\left( {1 - \mu _j^{k|k - 1}} \right)}}{{{{\left[ {\sigma _j^{k|k - 1}} \right]}^2}}} - 1} \right)\mu _j^{k|k - 1},	\\
	v_{S,j}^{k|k - 1} =& \left( {\frac{{\mu _j^{k|k - 1}\left( {1 - \mu _j^{k|k - 1}} \right)}}{{{{\left[ {\sigma _j^{k|k - 1}} \right]}^2}}} - 1} \right)\left( {1 - \mu _j^{k|k - 1}} \right),\\
	\mu _j^{k|k - 1} =& \frac{{u_j^{k - 1}}}{{u_j^{k - 1} + v_j^{k - 1}}},\\
	{\left[ {\sigma _j^{k|k - 1}} \right]^2} =& |{k_\beta }|\frac{{u_j^{k - 1}v_j^{k - 1}}}{{{{\left( {u_j^{k - 1} + v_j^{k - 1}} \right)}^2}\left( {u_j^{k - 1} + v_j^{k - 1} + 1} \right)}},\\
	\widehat m_{S,j}^{k|k - 1} =& \left[ {\widehat m_j^{k - 1};F \cdot \widehat m_{j,[k-1]}^{k - 1}} \right],\\
	\widehat P_{S,j}^{k|k - 1} =& \left[ {\begin{array}{*{20}{c}}
			{\widehat P_j^{k - 1}}&{\widehat P_{j,\left[ {t_j:k - 1,k - 1} \right]}^{k - 1}{F^\top}}\\
			{F\widehat P_{j,\left[ {k - 1,t_j:k - 1} \right]}^{k - 1}}&{F\widehat P_{j,[k-1,k-1]}^{k - 1}F^\top + Q}
	\end{array}} \right].
\end{align}
\par The prediction of trajectory in the BG-R-TPHD filter roots in the change of the target’s kinematic state. In contrast, the prediction of clutter only concerns the number of clutter generators in birth and surviving. The prediction of detection probability is completely governed by Beta densities. 
\begin{Pro}\label{BGUTPHD_Up}
	If at time $k$, the prior PHD $D_{k|k-1}(\tilde X)$ and $D_{k|k-1}(\tilde c)$
	are given and both PHD are given as Beta–Gaussian mixtures of the form
	\begin{align*}
		{D_{k|k - 1}}\left( {\tilde X} \right) =& \sum\limits_{j = 1}^{{J^{k|k - 1}}} {\omega _j^{k|k - 1}\beta \left( {a;u_j^{k|k - 1},v_j^{k|k - 1}} \right)}\notag\\
		&{\times{\cal N}\left( {X;t_j,\widehat m_j^{k|k - 1},\widehat P_j^{k|k - 1}} \right)},\\
		{D_{k|k - 1}}\left( {\tilde c} \right) =& \sum\limits_{j = 1}^{{J_c^{k|k - 1}}} {\omega _{c,j}^{k|k - 1}\beta \left( {o;u_{c,j}^{k|k - 1},v_{c,j}^{k|k - 1}} \right)},\notag
	\end{align*}
	then, given a measurement set $Z_k$, the posterior PHD $D_k(\tilde X)$ and $D_k(\tilde c)$ are given as
\end{Pro}
\begin{align}
	{D_k}\left( {\tilde X} \right)=& \sum\limits_{j = 1}^{{J^{k|k - 1}}} {\omega _j^{v,k}\beta } \left( {a;u_j^{k|k - 1},v_j^{k|k - 1} + 1} \right)\notag\\
	&\times{\cal N}\left( {X;t_j,\widehat m_j^{k|k - 1},\widehat P_j^{k|k - 1}} \right)\\
	&+ \sum\limits_{z \in {Z_k}} {\sum\limits_{j = 1}^{{J^{k|k - 1}}} {\omega _j^{u,k}(z)\beta } \left( {a;u_j^{k|k - 1} + 1,v_j^{k|k - 1}} \right)}\notag\\
	&\times{{\cal N}\left( {X;t_j,\widehat m_j^k\left( z \right),\widehat P_j^k} \right)}, \notag\\
	{D_k}\left( {\tilde c} \right)=& \sum\limits_{j = 1}^{{J^{k|k - 1}}} {\omega _{c,j}^{v,k}\beta } \left( {o;u_{c,j}^{k|k - 1},v_{c,j}^{k|k - 1} + 1} \right)\notag
	\end{align}	
	\begin{align}
	&+ \sum\limits_{z \in {Z_k}} {\sum\limits_{j = 1}^{{J^{k|k - 1}}} {\omega _{c,j}^{u,k}(z)\beta } \left( {o;u_{c,j}^{k|k - 1} + 1,v_{c,j}^{k|k - 1}} \right)},
\end{align}
where
\begin{align}
	\omega _j^{v,k} &= \omega _j^{k|k - 1}\psi^0[u_j^{k|k-1},u_j^{k|k-1}],\\
	\omega _j^{u,k} &=\notag\\
	&\frac{\omega _j^{k|k - 1}\psi^1[u_j^{k|k-1},u_j^{k|k-1}]{q_j}(z)}{\sum\limits_{l = 1}^{{J_c^{k|k - 1}}} {d^{k|k-1}_{c,l}\omega _{c,l}^{k|k - 1}\bar c(z)} + \sum\limits_{l = 1}^{{J^{k|k - 1}}} {d^{k|k-1}_{l}\omega _{l}^{k|k - 1}q_l(z)} },\\
	\omega _{c,j}^{v,k} &= \omega _{c,j}^{k|k - 1}\psi^0[u_{c,j}^{k|k-1},u_{c,j}^{k|k-1}],\\
	\omega _{c,j}^{u,k} &=\\
	& \frac{\omega _{c,j}^{k|k - 1}\psi^1[u_{c,j}^{k|k-1},u_{c,j}^{k|k-1}]\bar c(z)}{\sum\limits_{l = 1}^{{J_c^{k|k - 1}}} {d^{k|k-1}_{c,l}\omega _{c,l}^{k|k - 1}\bar c(z)} + \sum\limits_{l = 1}^{{J^{k|k - 1}}} {d^{k|k-1}_{l}\omega _{l}^{k|k - 1}q_l(z)} },\notag\\
	{\bar z _j} &= {H_j}\widehat m_{j,[k]}^{k|k - 1},\\
	{S_j} &= {H_j}\widehat P_{j,[k,k]}^{k|k - 1}H_j^\top + R,\\
	\widehat{P} _j^k &= \widehat{P}_j^{k|k - 1} - {K_j}{H_j}\widehat{P}_{j,\left[ {k,t_j:k} \right]}^{k|k - 1},\\
	{K_j} &= \widehat{P} _{j,\left[ {t_j:k,k} \right]}^{k|k - 1}H_j^\top S_j^{ - 1},\\
	\psi^0[u,v]&=\frac{B(u,v+1)}{B(u,v)},\\
	\psi^1[u,v]&=\frac{B(u+1,v)}{B(u,v)},\\
	d_{c,l}^{k|k-1}&=\frac{u_{c,l}^{k|k-1}}{u_{c,l}^{k|k-1}+v_{c,l}^{k|k-1}},\\
	d_{l}^{k|k-1}&=\frac{u_{l}^{k|k-1}}{u_{l}^{k|k-1}+v_{l}^{k|k-1}},
\end{align}
\par Different from traditional roust PHD filter \cite{UPD1}, the BG-R-TPHD filter aims at the whole trajectory in the update step. It not only updates the estimation of the target state at current time, but also smooths estimation of the previous states. 
\subsection{L-scan Approximation}
\par In this section, we will apply the $L$-scan approximation \cite{TPHD} to the BG-R-TPHD filter, which only considers the trajectory of the last $L$ time. If the length of trajectory $i\le L$, the prediction and update steps are the same as section IV.B. Otherwise, the prediction step changes into
\begin{align}
	\widehat{m} _j^{L,k|k - 1} =& \left[ {\widehat{m}_{j,\left[ {2:L} \right]}^{L,k - 1};F \cdot \widehat m_{j,[k-1]}^{k - 1}} \right],\\
	\widehat{P}_j^{L,k|k - 1} =& \left[ {\begin{array}{*{20}{c}}
			{\widehat{P}_{j,\left[ {2:L,2:L} \right]}^{L,k - 1}}&{\widehat{P}_{j,\left[ {2:L,L} \right]}^{L,k - 1}{F^\top}}\\
			{F\widehat{P} _{j,\left[ {L,2:L} \right]}^{L,k - 1}}&{F\widehat P_{j,[k-1,k-1]}^{k - 1}F^\top + Q}
	\end{array}} \right],
\end{align}
where the mean $\widehat{m}^{L,k-1}=[m^{k-L};...;m^{k-1}]\in \mathbb{R}^{Ln_{x}}$ and the covariance $\widehat{P}^L\in \mathbb{R}^{Ln_{x}\times Ln_{x}}$. The update step can be obtained by 
\begin{align}
	{\bar z _j} =& {H_j}\widehat m_{j,[k]}^{k|k - 1},\\
	{S_j} =& {H_j}\widehat P_{j,[k,k]}^{k|k - 1}H_j^\top + R,\\
	\widehat{m}_j^{L,k}\left( z \right) =& \widehat{m}_j^{L,k|k - 1} + {K_j}\left( {z - {{\bar z}_j}} \right),\\
	\widehat{P} _j^{L,k} =& \widehat{P}_j^{L,k|k - 1} - {K_j}{H_j}\widehat{P}_{j,\left[ {L,1:L} \right]}^{L,k|k - 1},\\
	{K_j} =& \widehat{P} _{j,\left[ {1:L,L} \right]}^{L,k|k - 1}H_j^\top S_j^{ - 1}.
\end{align}
\par Besides, we use a matrix $A_j^k=[m_j^{t_j};...;m_j^{k-L}]\in \mathbb{R}^{(i_j^k-L)n_x}$ to reserve the trajectory outside the $L$-scan window. The intact trajectory information consists of contents in the matrix $A$ and the corresponding $L$-scan window, which can be written as $\widehat{m}_j^k=[A_j^k;\widehat m_j^{L,k}]$. 
\subsection{Estimation}
For the BG-R-TPHD filter, the estimation of the number of alive trajectories at time $k$ is given as
$
	{N^k} =  \text{round}( {\sum_{j = 1}^{J^k} {\omega _j^k} }).
$
Then, the detection probability of the $j_{th}$ trajectory is ${p_{D,k,j}} =  {u_j^k}/(u_j^k + v_j^k)$ and the clutter rate is given as
\begin{equation}
	{N_c} =  \text{round}\left({\sum\limits_{j = 1}^{{J_c^k}} \frac{u_{c,j}^k}{u_{c,j}^k + v_{c,j}^k}\times{\omega _{c,j}^k} }\right).
\end{equation}
The estimations of $N^k$ trajectories are given as $
\left\{ {\left( {{t_1},i_1^k,\widehat m_1^k} \right),...,\left( {{t_{{N^k}}},i_{{N^k}}^k,\widehat m_{{N^k}}^k} \right)} \right\}.
$

\section{Simulations}
This section presents numerical studies for the BG-R-TPHD filter. We will do a four targets simulation inside of a two-dimensional space with the size of $\left[ {{\rm{ - 2000}},{\rm{2000}}} \right]m\times\left[ {{\rm{ 0}},{\rm{2000}}} \right]m$ for 100 seconds. The target state matrix is given as $x_k = {[p_k,\zeta_k ]^\top}$, where $p_k= [{p_x},{{\dot p}_x},{p_y},{{\dot p}_y}]$ denotes the position and velocity information and $\zeta_k$ denotes the turn rate information. The observation is given as
\begin{align*}
	z_k =\left[ {\begin{array}{*{20}{c}}
			{\text{arctan}\left(\frac{p_{x,k}}{p_{y,k}}\right)}\\
			{\sqrt{p^2_{x,k}+p^2_{y,k}}}
	\end{array}} \right]+v_k, 
\end{align*}
where $v_k\sim {\cal N}(\cdot;0,R_k),R_k=\text{diag}([(\pi/180)^2,4]^\top)$. The dynamic process for the single target is given as
\begin{align*}	
	p_{k+1}=&Fp_k+G\zeta_k,\\
	\zeta_{k+1}=&\zeta_k+\delta u_k,
\end{align*}
where
\begin{align*}	
	F =& \left[ {\begin{array}{*{20}{c}}
			{1}&{0}&{\frac{sin\zeta\delta t}{\zeta}}&{-\frac{1-cos\zeta\delta t}{\zeta}}\\
			{0}&{1}&{\frac{1-cos\zeta\delta t}{\zeta}}&{\frac{sin\zeta\delta t}{\zeta}}\\
			{0}&{0}&{cos\zeta\delta t}&{-sin\zeta\delta t}\\
			{0}&{0}&{sin\zeta\delta t}&{cos\zeta\delta t}
	\end{array}} \right],	G = \left[ {\begin{array}{*{20}{c}}
			{\frac{\delta t^2}{2}}&{0}\\
			{0}&{\frac{\delta t^2}{2}}\\
			{\delta t}&{0}\\
			{0}&{\delta t}
	\end{array}} \right],
\end{align*}
where $\delta t = 1$, $\zeta \sim {\cal N}(\cdot;0,\sigma^2_\zeta I_2)$, $\sigma^2_\zeta=1$m/$s^2$,$u_k\sim {\cal N}(\cdot;0,\sigma^2_u)$, $\sigma^2_u=(\pi/180)$rad/s. The surviving probability for target is given as ${p_{S}} =0.99$ and ${p^c_{S}} =0.9$ for clutter. The detection probability of targets is unknown for filter and given as ${p_{D}} =0.98$. The clutter process is of binomial distribution with $N_c =20$ and ${p^c_{D}} =0.5$, which is also unknown. Besides, the clutter is uniformly distributed in region $S=\left[ {{\rm{ - 2
			\pi}},{\rm{2\pi}}} \right]\times\left[ {{\rm{ 0}},{\rm{2000}}} \right]m$. The four targets initial states are given in TABLE \ref{Target States}
\par Besides, considering the Beta-Gaussian mixture model, we let $|k_\beta|=1.1$. The birth process is Poisson with parameters ${J_\gamma }=4$, ${\omega_\gamma }=0.01$, ${\widehat P_\gamma }=\text{diag}([50,50,50,50,3\pi/180]^2)$. For each birth $j \in \left\{ {1,2,3,4} \right\}$, states are given as $ \widehat m_{\gamma,1}^k=[-1500,250,0,0,0]^\top, \widehat m_{\gamma,2}^k=[-250,1000,0,0,0]^\top,  \widehat m_{\gamma,3}^k=[250,750,0,0,0]^\top, \widehat m_{\gamma,4}^k=[1000,1500,0,0,0]^\top$. The Beta distribution factors of birth are given as $u=8, v=2$ and $u_c=1, v_c=1$ for clutter.
\begin{table}[!t]
	\centering
	\caption{The Initial Target States}
	\label{Target States}
	\begin{tabular}{c|c|c|c}
		\hline
		&Kinematic State&Birth Time$/s$&Death Time$/s$\\
		\hline
		Target 1&$\left[1005,1489,8,-10\right]^{\top}$&1&100\\	 \hline
		Target 2&$\left[-256,1011,20,3\right]^{\top}$&10&100\\	 \hline
		Target 3&$\left[-1507,257,11,10\right]^{\top}$&10&80\\
		\hline 
		Target 4&$\left[250,750,-40,25\right]^{\top}$&40&100\\
		\hline
	\end{tabular}
\end{table}
\par In pruning and absorption, we use the a weight threshold of 
$\Gamma_p=10^{-5}$ and the threshold of absorption $\Gamma_a=4$ and the maximum of $J_{max}=100$. Besides, we use the trajectory metric error (TM) \cite{TM} with parameters $p = 2, c = 10, \gamma  = 1$ for the BG-R-TPHD filter. By running 1500 Monte Carlo, the performance of the BG-R-TPHD filter is given as follows. 
\begin{figure}[!t]
	\centering
	\includegraphics[width=3.5in]{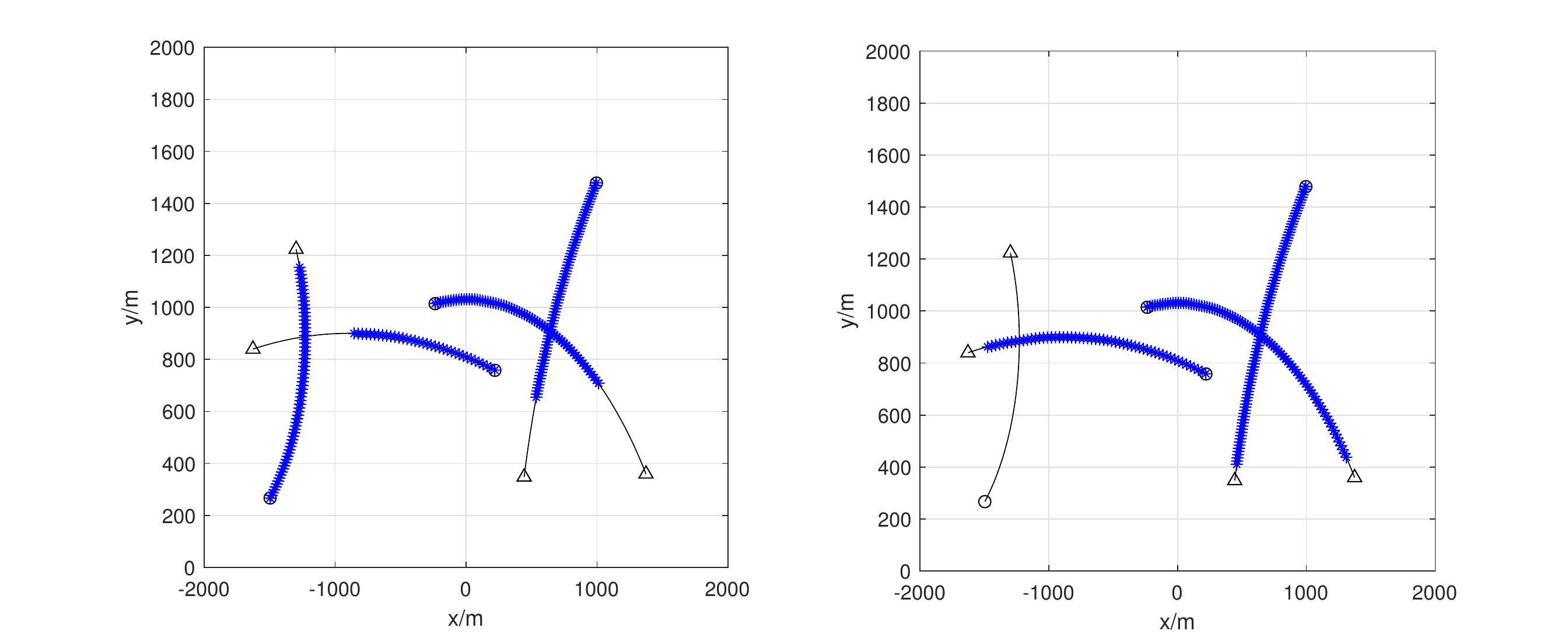}
	\caption{The trajectory of BG-R-TCPHD (left) at 75s (left), at 95s (right). Blue lines represent the estimation and black lines represent the truth. The circle represents the starting position of trajectories and the triangle represents the death position.}
	\label{states}
\end{figure} 
\begin{figure}[!t]
	\centering
	\includegraphics[width=2.8in]{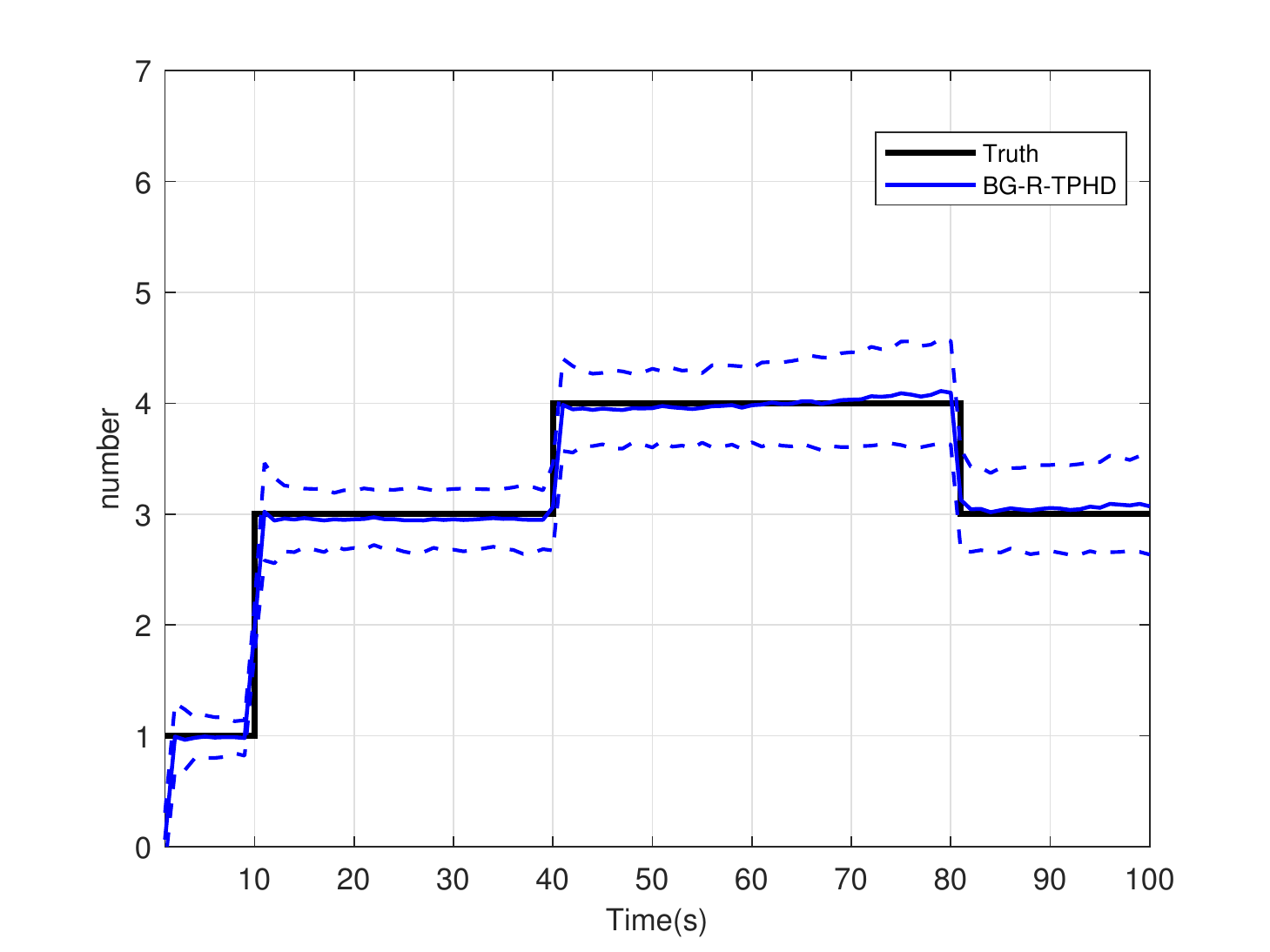}
	\caption{The estimation of the number of alive trajectories in the BG-R-TPHD filter. The real lines represent the number estimation, and the hidden lines represent estimation after calculating the standard deviation.}
	\label{numbers}
\end{figure}
\begin{figure}[!t]
	\centering
	\includegraphics[width=2.8in]{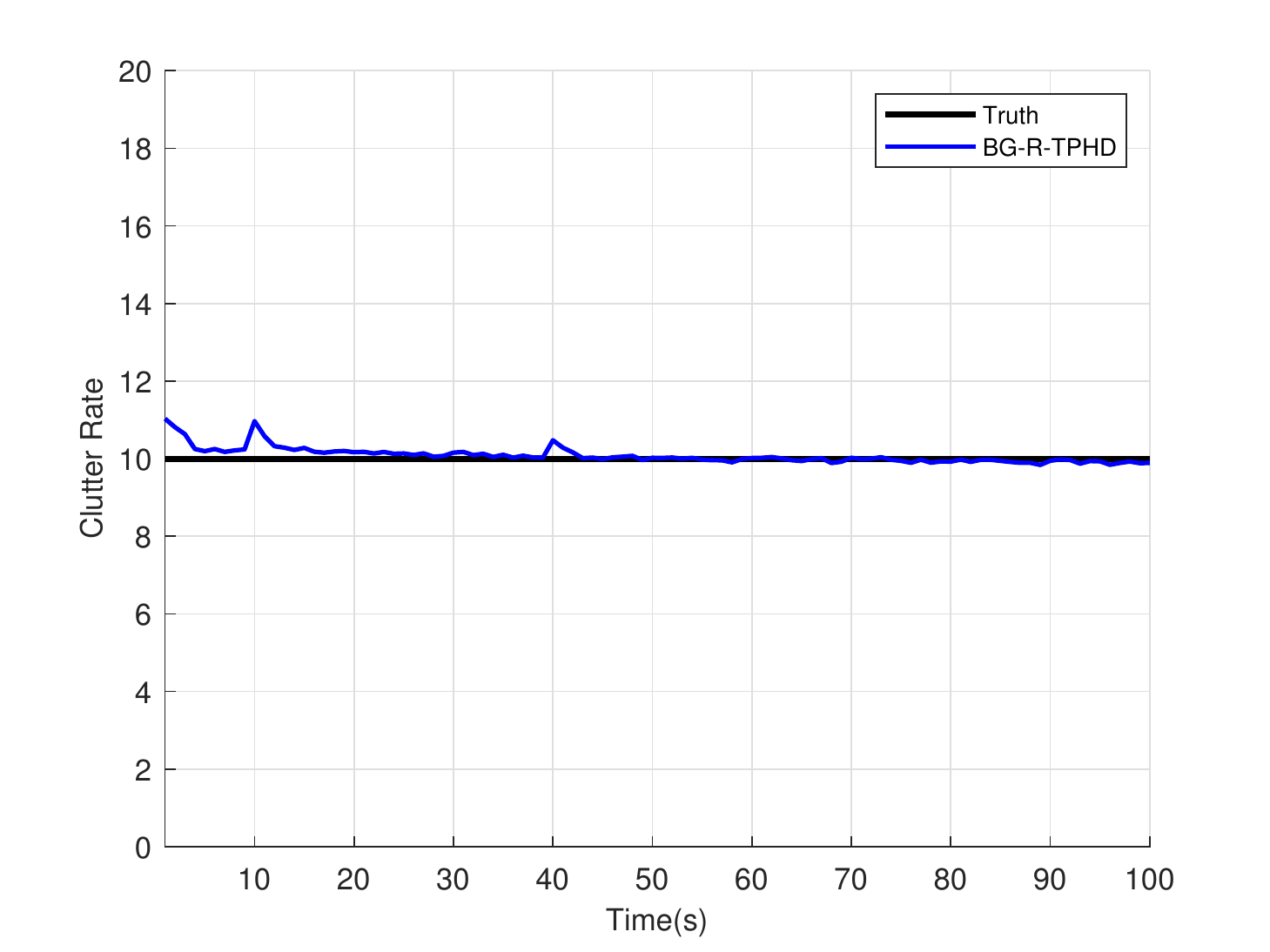}
	\caption{The estimation of the clutter rate in the BG-R-TPHD filter. The blue line represents the estimation, and the black line represents the truth.}
	\label{Clutter}
\end{figure}
\begin{figure}[!t]
	\centering
	\includegraphics[width=2.8in]{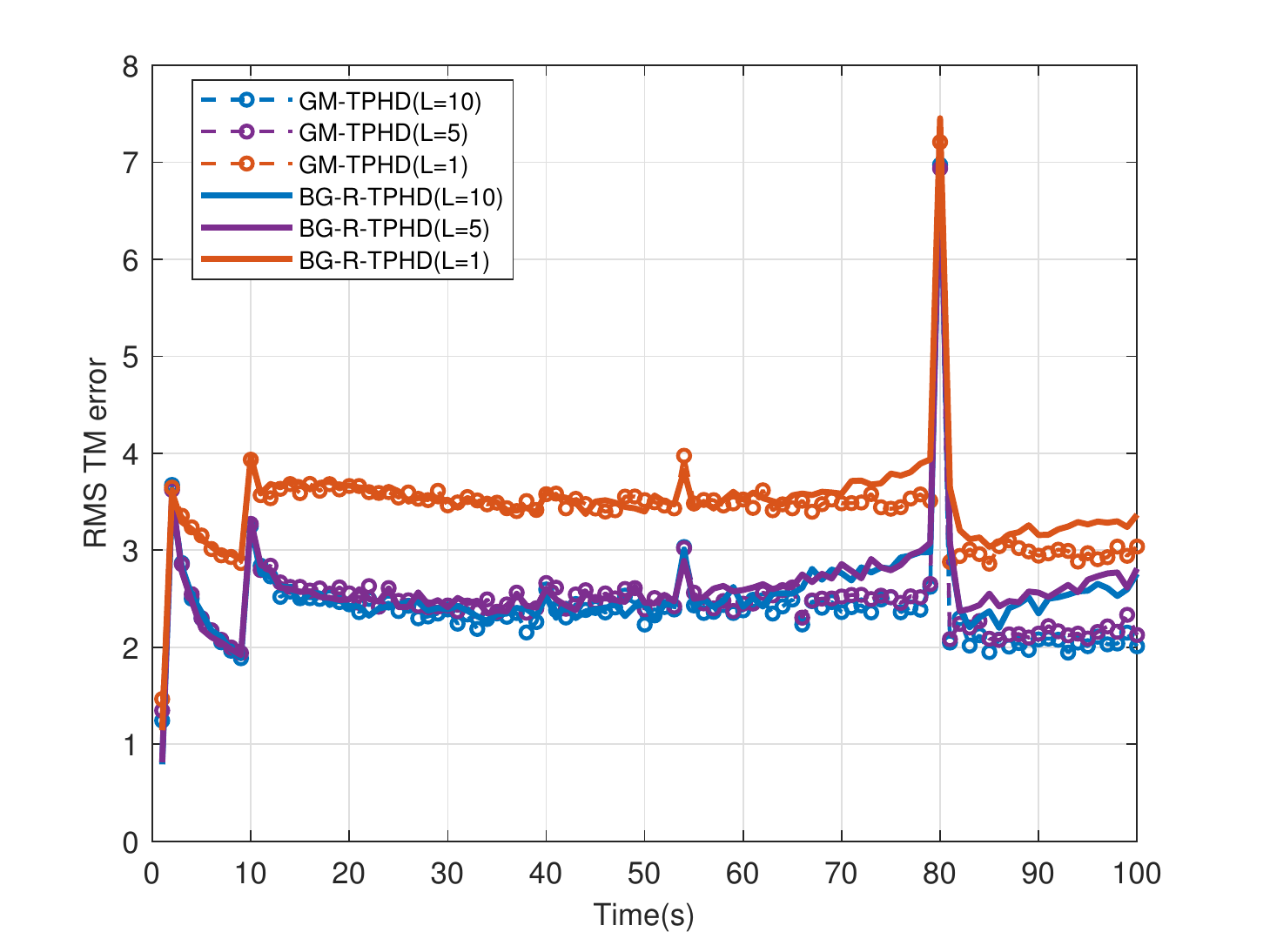}
	\caption{The RMS trajectory metric error for the BG-R-TPHD filter and GM-TPHD filter with known detection profile and clutter rate. Their performances with different $L$ are also shown.}
	\label{TM}
\end{figure}

\par It can be seen from Fig.\ref{numbers} and Fig.\ref{TM} that, the BG-R-TPHD filter can obtain excellent trajectory estimation which is close to the GM-TPHD filter with known detection profile and clutter rate. Besides, it can be seen from Fig.\ref{TM} that, usually a small value of $L$-scan can achieve almost full performance of the BG-R-TPHD filter with low computational burden and $L$=5 applies to our model. Besides, it can be seen from Fig.\ref{Clutter} that there is an initial settling in period for the estimation of clutter rate. However, after this miss distance, the estimation converges to the truth and keep stable but performs fluctuation with the birth and death of targets.
\begin{table}[!t]
	\centering
	\caption{The Run Time$(s)$ of The BG-R-TPHD Filter}
	\label{Lscan}
	\begin{tabular}{c|ccccccc}
		\hline
		&\multicolumn{7}{c}{Run Time$(s)$}\\
		\hline
		L&1&2&5&10&15&30&60\\
		\hline
		BG-R-TPHD&1.12&1.20&1.25&1.75&2.51&4.91&15.88\\
		\hline
	\end{tabular}
\end{table}

\section{Conclusion}
In this paper, we have derived the R-TPHD filter which can not only adaptively learn the unknown detection profile history and clutter rate, but also estimate trajectories, based on KLD minimization. Meanwhile, we also give another analytic form of the R-TPHD filter with unknown detection profile of a single frame as an approximation. We have proposed a suboptimal implementation for the R-TPHD filter based on Beta-Gaussian mixture model and their $L$-scan approximations are also proposed to achieve a fast implementation. Based on simulation results, the R-TPHD filter can achieve excellent performance close to the GM-TPHD filter with known detection probability and clutter rate.

\end{document}